# Aerodynamic Significance of Mass Distribution on Samara Descent

Zhao-Bang Hou[1,†], Jun-Duo Zhang[2,†], Yun-Da Li[1], Yong-Xia Jia[2], Wei-Xi Huang[2]*

[1] Xingjian College, Tsinghua University, Beijing 100084, China

[2] AML, Department of Engineering Mechanics, Tsinghua University, Beijing 100084, China

[†] These authors contributed equally to this work.

* Corresponding author: hwx@tsinghua.edu.cn

## Abstract

Samaras, a distinct category of fruit, are composed of heavier seeds and lighter wings. Diversity in morphologies and structures subtly contributes to the flight patterns of various seeds, thereby serving as a key factor in the reproductive strategies of plants. To explore the mechanisms underlying various samara flight behaviors, we proposed an effective scheme by manipulating the mass distribution on a plate to mimic various three-dimensional descent behaviors of samaras. Through this framework, we experimentally identified and characterized four distinct flight modes. The three-dimensional vortical structures were then numerically analyzed to gain insights into the samara-inspired flight behaviors. Our study demonstrates how strategic mass distribution in samaras leads to diverse flight behaviors that leverage vortices to enhance seed dispersal, offering a fresh perspective for the design of biomimetic fliers.

## Introduction

Seed dispersal is a fundamental ecological process that enhances plant reproduction and drives evolutionary developments across diverse species[1, 2, 3, 4]. Samaras, such as those from maple, ash, elm, and mahogany trees, consist of heavier seeds with lighter wings. This unique configuration enables them to descend slowly and exploit wind currents, facilitating seed dispersal over distances ranging from several meters to kilometers[5, 6, 7]. In pursuit of this, a variety of morphologies and structural adaptations have evolved (see Figure 1A), resulting in distinct flight patterns that leverage aerodynamics in different ways.

The flight capabilities of samaras have long intrigued researchers across various fields. Studies have documented a wide range of flight behaviors among different samara species, suggesting a correlation between these behaviors and the samaras' mass distribution[8, 9, 10]. For example, samaras from maple and mahogany trees, with mass concentrated at the terminal end and leading edge, exhibit autorotative flight[11, 12, 13] (Figure 1B). Research on maple samaras shows that this specific mass distribution facilitates the formation of the leading edge vortex (LEV) to generate lift[8, 14, 15, 16, 17], a mechanism also observed in animal locomotion to enhance propulsion[18, 19, 20, 21, 22]. Alternatively, samaras from ash and elm trees, which have more centrally distributed mass, tend to tumble downward along curved trajectories[12, 23, 24, 25] (Figure 1B). Additionally, *Ventilago leiocarpa* samaras, with a heavy seed positioned at one end of the wing, display a rapid downward flight pattern. Despite the structural similarities among these samaras, the subtle variations in their mass distribution may be linked to the diversity of flight modes they exhibit, representing a significantly underexplored area of research.

Freely falling bodies are a common occurrence in nature[26]. Previous studies have primarily focused on the two-dimensional (2D) descent of plates, showing that variations in rotational inertia and Reynolds number can lead to transitions between tumbling, fluttering, and chaotic trajectories[27, 28, 29, 30]. However, the vortical structures and dynamics associated with three-dimensional (3D) falling trajectories differ remarkably[31, 32]. For instance, a circular disk undergoes helical spiral descents, with flow structures significantly different from those observed in 2D motions[33, 34, 35]. Given the diverse structural configurations and flight patterns of samaras, exploring their behaviors in 3D scenarios is essential.

Since samaras facilitate flight solely through structural configuration without the neuromuscular control present in animals[17], they have provided significant inspiration for the field of robotics. Electronical samara-inspired fliers have been designed for environmental monitoring and other applications[36, 37, 38, 39]. A thorough understanding of the fundamental mechanisms governing the flight patterns of samaras is vital for designing these bioinspired devices. Existing designs typically adopt a single configuration to achieve specific flight behaviors, suggesting that investigating more diverse and flexible control strategies could be beneficial.

In this study, we introduced a novel framework to explore a broad range of samara-inspired flight behaviors. By manipulating the mass distribution on a plate to mimic the descending behaviors of various samaras, we experimentally identified four distinct flight modes and analyzed their flight patterns. Using computational fluid dynamics (CFD), we focused on the characteristic vortices underlying 3D flight patterns. Our research highlights the critical role of mass distribution in the 3D descent of samaras and its influence on the formation of vortices and wake interactions that produce lift and sustain flight patterns. These findings offer potential strategies for designing future flying devices.

## Results

### Samara-inspired framework

Focusing on the influence of seed placement on aerodynamic behaviours of samaras, we developed a framework that enables manipulation of mass distribution to emulate various falling patterns inspired by wing-seed structures (Figure 1C). Observations indicate samaras commonly possess wing aspect ratios ranging from 3 to 5 to enhance their flight stability[40, 41], and thus we employed a rectangular thin plate with an aspect ratio of 3:1 to represent the typical lightweight flattened wing structure of samaras. To mimic the dense nut and the unevenly distributed mass on the wings of real samaras, two weights were attached to the plate: a heavier weight ($m_h$) placed along the longer axis, and a lighter weight ($m_l$) placed along the shorter axis. By adjusting the distance of $m_h$ and $m_l$ from the plate's centerlines (denoted as $x'$ and $y'$, respectively), we can precisely control the mass distribution, thus altering the center of mass (COM) location $(x_c, y_c)$, which is calculated as follows:

$$x_c = m_h \cdot x'/m_{total} \qquad (1-1)$$

$$y_c = m_l \cdot y'/m_{total} \qquad (1-2)$$

## Identification of distinct flight modes

By adjusting the weight position on the plate, we identified four distinct flight modes: Autorotation (AR), Spiral Tumbling (ST), Chaotic (CH), and Falling (FA). These modes are systematically mapped in Figure 2A across the parameter space defined by the COM location, with their corresponding flight patterns illustrated in Figure 2B-2F. Additionally, our measurements of natural samaras' mass distribution align well with the results from the samara-inspired framework (see Supplementary Information).

The AR mode exhibits pronounced periodicity, triggered when both weights are placed towards the plate's edges. This mode represents samaras with autorotative flight patterns, such as those of maple and mahogany trees. These samaras from different groups of plants have convergently evolved into a configuration where heavier seed is positioned near the terminal end and a lighter wing with ridged leading edge[42, 43]. This configuration facilitates stable descent while rotating around a vertical axis (Figure 2B), enabling efficient and stable dispersal[11, 12, 13].

The ST mode primarily features the plate rapidly tumbling during spiraling descent. Within this mode, two divergent spatial trajectories emerge – continuous and segmented ST motions (see Figure 2C and 2D, respectively). The continuous ST flight follows a steady helical path around a vertical axis. This continuous descent motion represents the dispersal strategies of several natural samaras, such as those from elm and ash trees, which utilize a similar flight pattern for seed propagation[12, 23, 24, 25]. This continuous flight pattern is achieved when both weights are centrally placed, mimicking the structure of samaras with seeds or mass concentrated near the wing's center. In contrast, the segmented ST flight presents a distinct trajectory with multiple turning points between each spiral phase. This variation arises when the lighter weight is positioned further from the plate's centerline, resulting in a unique spatial trajectory.

Located between the ST and AR modes on the parameter map, the CH mode is characterized by irregular rotational and translational flight patterns during descent (Figure 2E). This mode exhibits chaotic behavior, similar to the motion observed in the falling of disks and plates, characterized by random transitions between tumbling and fluttering motions[30, 35, 44]. The CH mode lacks consistent motion patterns and demonstrates substantial oscillations in both velocity and orientation, making the flight motion appear unpredictable to the observer.

The FA mode occurs when the heavier weight is near the terminal end and the lighter weight is positioned close to the plate's centerline, resembling the mass distribution found in *Ventilago leiocarpa* samaras. This configuration intuitively leads to a rapid descent, with the plate tilting towards the side with the concentrated mass, thereby significantly increasing the falling speed (Figure 2F).

## Kinematics of typical flight patterns

To quantitatively examine the kinematic behaviors of typical flight patterns, we measured both the translational and rotational motions of the plate during descent. Utilizing six-degree-of-freedom kinematic data, the 3D descent trajectories for various flight modes are reconstructed.

In the AR mode ($x_c/a = 0.39$, $y_c/b = 0.17$), the trajectory mainly exhibits a vertical descent with slight horizontal displacement. The 3D reconstruction reveals that the plate rotates around the vertical axis while maintaining a small cone angle ($\theta = -11.4 \pm 4.2°$) and virtually no tilt ($\psi \approx 0°$) (Figure 3A).

The ST mode primarily involves tumbling motion, with variations in mass distribution leading to distinct spatial trajectories. In the continuous ST flight ($x_c/a = 0.06$, $y_c/b = 0.04$), the plate follows a steady helical path, maintaining a nearly constant cone angle ($\theta = -38.2 \pm 2.3°$) between its longer axis and the horizontal plane (Figure 3B). In contrast, the segmented ST flight ($x_c/a = 0.10$, $y_c/b = 0.10$) exhibits dynamics similar to the continuous ST motion but with more pronounced fluctuations in the cone angle ($\theta = -38.3 \pm 13.0°$). This trajectory features abrupt transitions at turning points, where both translational and rotational movements change drastically (Figure 3C). The divergence between the continuous and segmented ST trajectories is likely due to mass distribution asymmetry. In 2D scenarios, plates with symmetric mass distributions tend to exhibit tumbling motion toward a single direction. However, asymmetric front-weighting results in tumbling modes that involve flips in their trajectory[45]. This may also explain why, in 3D situations, a shift in mass distribution away from symmetry drives the transition from continuous to segmented spiral tumbling motion.

To compare the continuous and segmented ST trajectories, we plotted their top-view projections in Figure 3D and 3E, respectively. In the continuous ST motion ($x_c/a = 0.06$, $y_c/b = 0.04$) (Figure 3D), the plate performs a counterclockwise circular motion in the horizontal plane with a radius of approximately $1.25L$. During its descent, the plate completes roughly seven tumbling rotations per spiral revolution, as its tumbling angular velocity significantly exceeds that of its spiral motion. Conversely, the segmented ST motion ($x_c/a = 0.10$, $y_c/b = 0.10$) displays a trajectory featuring a critical turning point (Figure 3E, left). Initially, the motion resembles counterclockwise ST, which is then shifted to clockwise ST after the turn. The horizontal projection of both segments is close to circular segments with a radius of about $1.89L$. As the plate approaches the turning point, its cone angle increases, followed by a reversal in the tumbling direction. A zoomed-in view of the trajectory around the turning point (Figure 3E, right) reveals a peak self-rotation angle ($\psi \approx 0.4\pi$), accompanied by a significant angular reversal. Compared to the continuous ST motion, the sudden transition introduces significantly higher fluctuations in angular velocities (Figure 3F), highlighting the unique dynamic transitions in the segmented ST motion.

In the CH mode ($x_c/a = 0.18$, $y_c/b = 0.08$) (Figure S1), we observed transitions between two primary forms of motion. The plate may display periods of translational motion, where lateral drift and stable descent are prominent, or it may enter phases dominated by tumbling motion, with significant rotational motion and reduced horizontal drift. This mode exhibits unpredictable behavior, with irregular shifts between phases dominated by translational and rotational motions.

The FA mode ($x_c/a = 0.35$, $y_c/b = 0.08$) is characterized by a rapid descent with limited horizontal displacement (Figure S2). Throughout most of the motion, the plate maintains a relatively large and stable tilting angle, with the heavier side consistently facing downward. This results in a relatively small windward area, thus increasing the falling speed.

## The statistical characteristics of descent process

To explore the dispersal capabilities of various flight modes in quiescent air, we focused on the falling velocity and horizontal propagation distance of static releases. We selected typical configurations for each mode and conducted multiple releases from a fixed height and location. The landing point and time of each release were recorded, allowing us to calculate the average falling velocity ($V_d$) and dimensionless horizontal propagation range ($R/H$), as depicted in Figure 4.

Although the AR mode shows a limited horizontal propagation range in quiescent air, it benefits from the lowest falling velocity, resulting in an extended duration aloft. This mode's efficient lift generation is particularly advantageous in the presence of crosswinds, enabling broader seed dispersal.

Conversely, the ST mode, despite its faster falling velocity compared to the AR mode, achieves a wider horizontal spread in quiescent air. This characteristic may explain why various samaras employ spiral tumbling mode to enhance seed dispersal. Specifically, while the horizontal propagation ranges of the continuous and segmented ST motions are comparable, the segmented ST motion demonstrates a higher average falling velocity. The greater cone angle near turning points likely reduces the windward area, leading to a faster descent.

Arising from its unpredictable flight dynamics, the CH mode showcases significant variability in both falling velocity and horizontal dispersion. Finally, the FA mode achieves the highest falling velocity and a relatively moderate horizontal propagation range, emphasizing its rapid vertical and slightly inclined descent.

## Aerodynamics and vortical structures

Through CFD simulations of the AR and ST modes, we explored the 3D vortical structures associated with samara-inspired falling trajectories. Our results reveal that leading-edge vortices (LEVs), trailing-edge vortices (TEVs), and tip vortices (TVs) are essential elements, shaping the aerodynamic behaviors of the samaras' falling trajectories in distinct ways.

Specifically, in the AR mode ($x_c/a = 0.35$, $y_c/b = 0.18$), the instantaneous 3D vortical structures at two different instants ($t_{A1}$ and $t_{A2}$) are visualized by the iso-surfaces of $Q$-criterion[46], as shown in Figure 5A. At $t_{A1}$, a prominent LEV forms as the rotating plate encounters the flow, inducing a strong negative pressure zone near the leading edge. By $t_{A2}$, the pressure distribution remains largely unchanged, indicating that the LEV stays stably attached to the plate and continues to contribute to lift. This force-enhancing mechanism, widely utilized in animal propulsion, also plays a vital role in samaras' flight[18, 19, 20, 21, 22]. Additionally, a TEV is observed along the plate's trailing edge, similarly generating lift by creating a low-pressure region. The TEV mechanism, which is observed in both insect flight[47, 48] and fish swimming[49, 50], illustrates shared aerodynamic strategies across different forms of life. Unlike the continuous tip vortex shedding observed in previous studies, several discrete rib-like TVs (RLVs) form at the plate's tip and are shed into the wake. They induce a downwash momentum that contributes to lift production[51, 52, 53, 54].

Instantaneous vortices in the ST mode ($x_c/a = 0.06$, $y_c/b = 0.00$) at various instants are demonstrated in Figure 5B. Due to the plate's tumbling motion around the plate's longer axis, the flow separates along the longer edge, forming a pronounced LEV that creates a low-pressure zone at $t_{S1}$. By $t_{S2}$ (valley of the lift force), the LEV sheds into the wake and no longer contributes to lift. During this tumbling motion, tip vortices (TVa, TVb) shed from two shorter edges and gradually stretch as the plate descends ($t_{S2}$). By $t_{S3}$ (peak of lift force), TVa and TVb are further elongated and eventually merge into a Ω-shaped vortex tube (ΩVa). The TEV, initially aligned with the trailing edge shortly after shedding ($t_{S1}$), gradually deforms under the induction of the stretching TVs and also forms an Ω-shaped vortex tube (ΩVb) by $t_{S3}$. These Ω-shaped vortex tubes generate a downwash momentum aligned with the tumbling direction, helping to sustain the plate's continuous rotation.

The projected forces and torques during a rotation cycle in the AR and ST modes are evaluated in Figure 5C. In the AR mode, all force and torque components exhibit only slight fluctuations. The vertical aerodynamic force (lift) is notably larger than the horizontal force components and the horizontal component of aerodynamic torque ($M_y$) dominates during descent, sustaining the revolving motion. In contrast, the aerodynamic properties in the ST mode display significant periodic fluctuations, corresponding to the tumbling motion. The vertical force ($F_z$) shows the largest oscillation magnitude among all components, and the longitudinal torque ($M_x$) possesses the greatest amplitude due to the dominance of the tumbling motion. Our theoretical analysis (see Supplementary Information) suggests that these differences in aerodynamic characteristics arise from the displacement between the center of lift and the COM, driven by the variation of mass distribution.

## Discussion

In this study, we proposed a scheme that manipulates mass distribution on a plate to present the diverse aerodynamic behaviors of samaras' flight. Through our framework, four distinct flight modes were identified: Autorotation (AR), Spiral Tumbling (ST), Chaotic (CH), and Falling (FA) (Figure 2A). Our measurements showed that the mass distributions of various natural samaras, which exhibit these flight modes, closely align with the present samara-inspired framework (see Supplementary Information). By reconstructing the trajectories of typical flight patterns using experimental kinematic data, we provided a quantitative analysis of their aerodynamic characteristics. Notably, our results revealed that variations in the COM could lead to transitions between periodic and chaotic flight patterns. Data on descent velocities and horizontal propagation distances demonstrated that the periodic modes like AR and ST exhibit relatively stable flight performance, suggesting a potential reason why such strategies are prevalent in natural samara dispersal.

We also conducted numerical simulations to understand the 3D vortical structures associated with samara-inspired falling. The analysis of flow structures and pressure distributions highlighted the critical roles of LEVs, TEVs, and TVs in generating lift and torque. Interesting rib-like TVs in the AR mode and intertwined Ω-shaped vortices in the ST mode were observed, both of which contribute to maintaining their respective flight patterns. Additionally, our theoretical analysis (see Supplementary Information) suggests that the aerodynamic differences between the AR and ST modes are due to the displacement between the center of lift and the COM.

Our findings revealed the impact of mass distribution on samaras flight patterns and vortical structures, thereby enhancing their aerodynamic performance. These findings suggest that by strategically manipulating the mass distribution, it is feasible to alter the flight patterns of samara-inspired aerial devices, potentially improving their maneuverability and reliability. This flight pattern control paradigm could be integrated with several established microflier control techniques, such as origami structures[37] and 3D electrothermal actuators[39], offering a new perspective for the design of biomimetic microfliers.

While our study primarily focused on the role of mass distribution in samara flight dynamics, factors such as surface texture[55] and 3D curvature[56, 57] were not considered. Nevertheless, the key insights into how mass distribution influences flight patterns and vortical structures remain well-supported. The approach proposed here can be adapted and modified to study a broader range of 3D samara flight behaviors, laying a foundation for future explorations into the complex flight dynamics of samaras.

## Methods

### Samara-inspired framework fabrication

We designed our model to mimic the aerodynamic structure of samaras using a rectangular thin plate with an aspect ratio of 3:1. This plate represents the lightweight, flattened wing structure typical of samaras. The plate, measuring $60mm$ by $20mm$ with a surface density of $77.5g/m^2$ for a total mass of $92.8mg$, was precisely cut from a kraft paper sheet. To mimic the mass concentration at the nut and leading edge of samaras, two metallic weights were attached to the plate: a heavier weight ($102.9mg$) along the longitudinal axis and a lighter weight ($51.4mg$) along the transverse axis. These weights, adjustable in position, allow for accurate control over the COM of the plate. During free-release experiments, these weights remain fixed on the plate, ensuring consistent mass distribution throughout each test.

### Experimental Setup

Experiments were conducted inside an enclosed 4m-height platform with a $2m \times 2m$ cross-section, under controlled environmental conditions of $15 \pm 1°C$ and standardized air density (Figure S3). The flight dynamics of the plate were captured using two high-speed cameras (Phantom VEO-640L) positioned at the top of the platform, recording at 1000 frames per second. A remotely controlled servo motor clamp was used for the plate release mechanism, ensured consistent release conditions (cone angle $\theta = 0$ and self-rotation angle $\psi = \pi/2$) in different test cases.

## Kinematics measurement technique

The motion of the falling plate was tracked using a stereoscopic vision system[33, 58, 59], which recorded the plate's trajectory via two synchronized cameras. The system was calibrated using a chessboard pattern, enabling accurate 3D spatial reconstruction. Key reference points on the plate were identified and tracked across frames to dynamically determine its position in real-time (Figure S4). The Levenberg-Marquardt algorithm was utilized to optimize the six degrees of freedom (translational and rotational), significantly reducing re-projection errors and enhancing the accuracy of motion analysis[33, 58, 59]. In instances where the plate moves perpendicularly to the camera's field of view, a second-order interpolation of motion parameters was performed to reconstruct the plate's trajectory and flight attitude effectively.

## Simulations

CFD simulations were conducted using the immersed boundary (IB) method, a technique widely applied in biomimetic flow studies[60, 61, 62, 63]. The Navier-Stokes equations for incompressible Newtonian flow are solved at the Reynolds number $Re = Lu_0/\nu = 560$ with the characteristic velocity $u_0 = \sqrt{gL}$. The IB method defines the flow field on a background Eulerian grid and the immersed boundary (i.e., the plate) on a Lagrangian mesh to facilitate efficient resolution of the unsteady flow-structure interaction. The plate is modeled as a rigid thin rectangle with dimensions $L \times L/3$, falling under gravity and aerodynamic forces with six degrees of freedom. The computational domain with a size of $4L \times 4L \times 4L$ was dynamically adapted to follow and cover the plate's extensive trajectories. The plate was discretized using a rectangular mesh with a resolution of $h/L = 0.0039$, while the overall computational domain was resolved on a uniform Cartesian grid with a grid number of $256 \times 256 \times 256$. The time step was set as $\Delta t = 0.00005 \times L/u_0$ to ensure the numerical stability and accuracy of capturing the plate's dynamics. The computational settings were chosen after testing for independence of domain size, grid resolution, and time step. All results presented were obtained after a duration of $10 \times L/u_0$ when the plate reaches a periodically steady state. The forces and torques presented in this study are all nondimensionalized by $\rho_f u_0^2 L^2$ and $\rho_f u_0^2 L^3$, respectively (Figure S5).

# Figures

**Figure 1: Framework inspired by samaras**

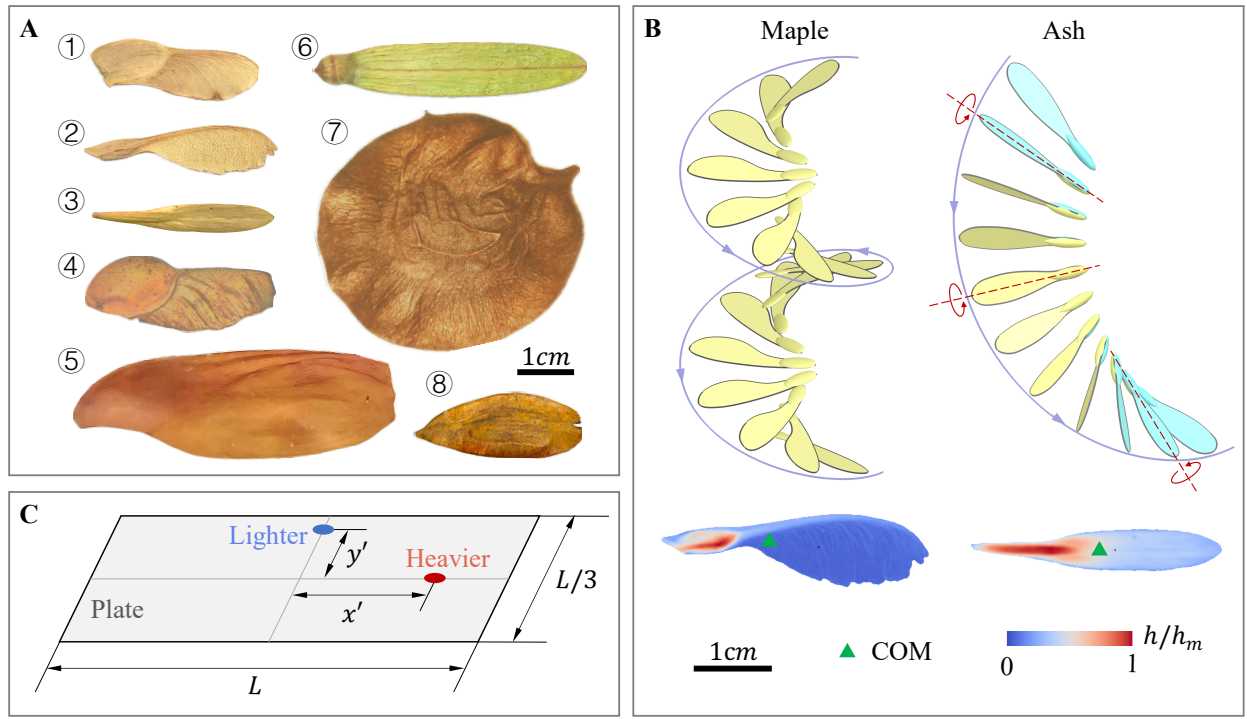

**(A)** Photographs of various types of samaras: ① *Acer pictum* (maple), ② *Acer negundo* (maple), ③ *Fraxinus* (ash), ④ *Pterolobium punctatum*, ⑤ *Swietenia mahagoni* (mahogany), ⑥ *Ventilago leiocarpa*, ⑦ *Pterocarpus santalinus* (red sandalwood), and ⑧ the *Eucommia ulmoides*. **(B)** Schematic of falling trajectories and mass distribution of maple and ash samaras. The normalized thickness distributions and center of mass (COM) locations are presented, based on data obtained from actual samaras. Maple samaras demonstrate autorotative flights due to mass concentration at the terminal end, whereas ash samaras, with more centrally distributed mass, spiral downward, rotating along their longitudinal axis. $h/h_m$ represents the normalized thickness. **(C)** Schematic of the samara-inspired framework. A rectangular thin plate with an aspect ratio of 3:1 mimics the typical light flattened wing structure of samaras. Heavier and lighter weights are attached to the plate to emulate the mass concentrations found at the nut and leading edge, respectively. Adjusting their distance from the plate's centerlines allows for precise control of the mass distribution.

**Figure 2: Parameter space and trajectories of distinct flight modes.**

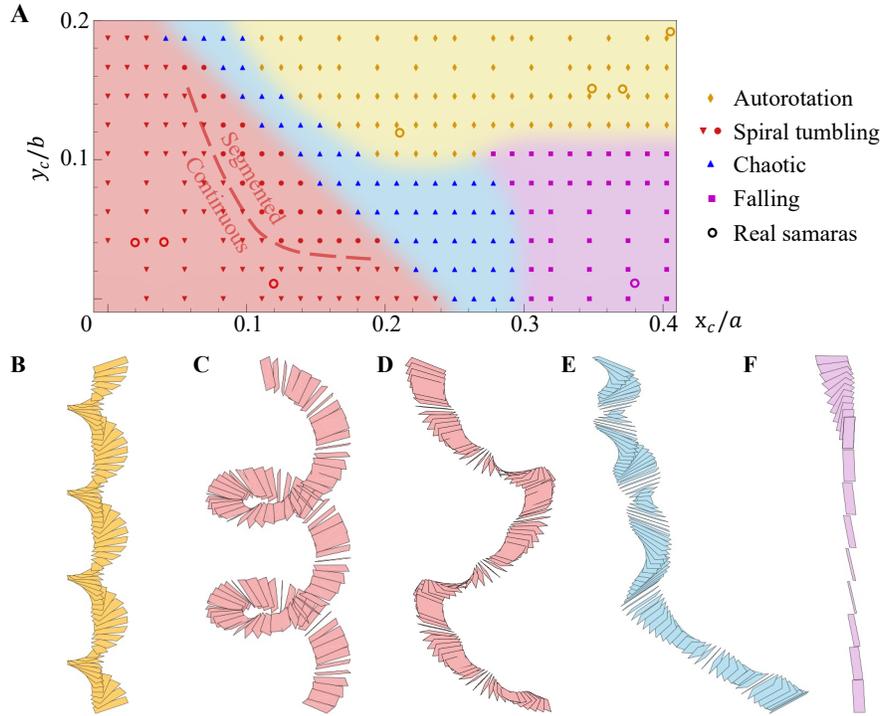

**(A)** Distribution of flight modes within the parameter space defined by the COM location. $x_c$ and $y_c$ represent the COM locations along the plate's longer and shorter axes, respectively, while $a$ and $b$ denote the half-lengths of the plate's longer and shorter axes, respectively ($a = L/2, b = L/6$). Yellow diamond scatters indicate mass configurations for the Autorotation (AR) mode, red triangles and circles for the continuous and segmented Spiral Tumbling (ST) modes respectively, blue triangles for the Chaotic (CH) mode, and purple squares for the Falling (FA) mode. Hollow circle scatters represent mass configurations of real samaras exhibiting these modes (see Supplementary Information for measurement details). **(B)** AR mode, characterized by stable descent with rotation around a vertical axis. **(C)** Continuous ST mode, featuring rapid tumbling and spiraling downward along a vertical axis. **(D)** Segmented ST mode, exhibiting rapid tumbling with distinct turning points in its trajectory. **(E)** CH mode, marked by irregular tumbling and oscillations with abrupt changes in direction. **(F)** FA mode, involving a rapid, predominantly vertical descent.

**Figure 3: Reconstruction of descent trajectories using experimental kinematic data.**

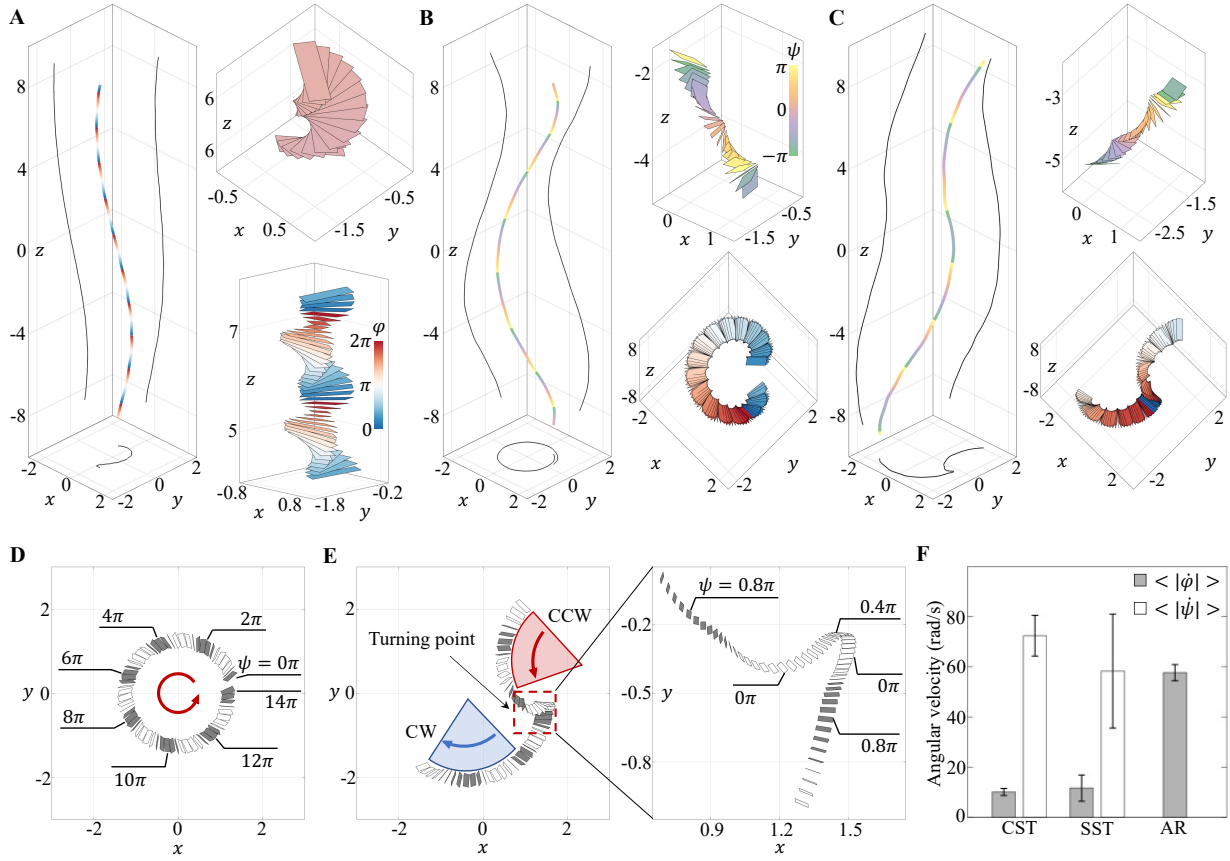

Trajectories of various flight modes are visualized, with data normalized by the plate's length ($L$) and reconstructions overlaid at $5ms$ intervals (see also movie S1). **(A)** AR mode. **(B)** Continuous ST mode. **(C)** Segmented ST mode. The 3D trajectories of the plate are shown along with its projections. 3D reconstructions in the upper right use the self-rotation angle ($\psi$) around the plate's longer axis, coloring the plate from green to orange. 3D reconstructions in the lower right use the revolution angle ($\varphi$) around the vertical axis, coloring the plate from blue to red. **(D)** Top view projection of the continuous ST motion. The variation of $\psi$ in a counterclockwise (CCW) revolution cycle is presented. **(E)** Top view projection of the segmented ST motion with a zoomed-in view near the turning point. The motion initially resembles the CCW ST, which is shifted to clockwise (CW) ST after the turn. The variation of $\psi$ near the turning point is displayed. **(F)** Comparison of the mean angular velocities ($<|\varphi'|>$ and $<|\psi'|>$) among the continuous ST (CST), segmented ST (SST), and AR modes, with ticks indicating the magnitude of fluctuation around the mean values.

**Figure 4: Distribution of average descent velocities ($V_d$) and horizontal propagation ranges ($R/H$) in quiescent air.**

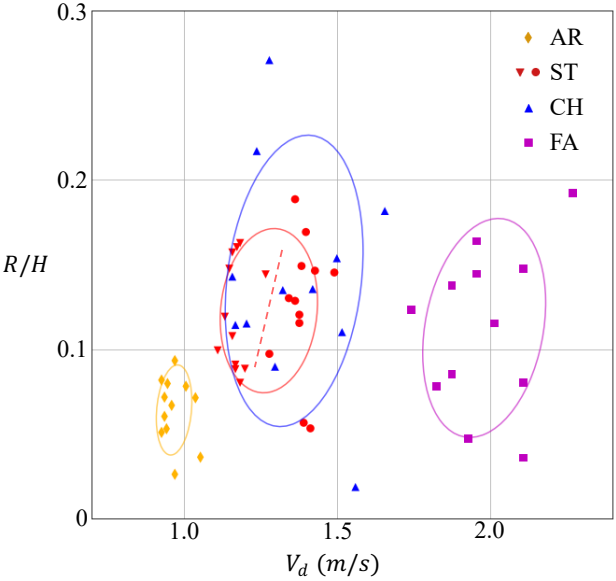

The graph displays $V_d$, the averaged vertical velocity, and $R/H$, the ratio of propagation distance to descent height. Yellow diamond scatters indicate mass configurations for the AR mode, red triangles and circles for the continuous and segmented ST modes respectively, blue triangles for the CH mode, and purple squares for the FA mode. Covariance ellipses with a 0.6 confidence level illustrate the statistical distribution of the different flight modes.

**Figure 5: Vortical structures and aerodynamic properties in periodic flight modes.**

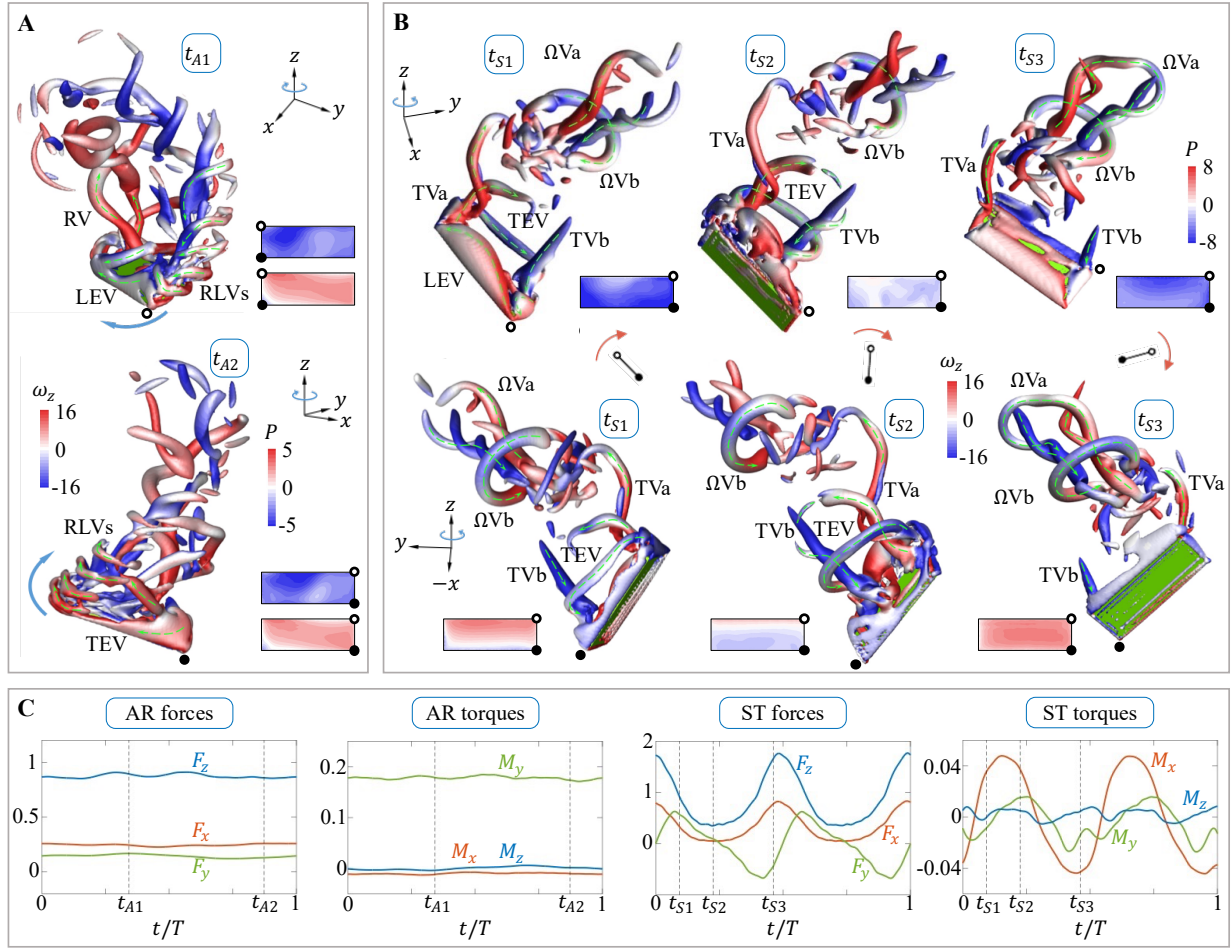

Iso-surface of $Q$-criterion ($Q = 60$) illustrates the vortical structures around the plate, colored by vertical vorticity ($\omega_z$). Pressure distributions on upper and lower surfaces are also depicted at several instants in the AR **(A)** and ST **(B)** modes, respectively. For each instant, the pressure distribution on the upper side of the plate is shown above, while the lower side is shown below. Orientation markers, represented by hollow and solid dots along the plate's shorter edge, indicate the plate's orientation. The vortical structures of the AR and ST modes are also shown in movie S2. **(C)** Evaluation of aerodynamic forces and torques during a motion cycle in the AR and ST modes. All force and torque components are projected as follows: the $x$-axis is defined by the projection of the plate's longer axis onto the horizontal plane, the $y$-axis is perpendicular to the $x$-axis within that plane, and the $z$-axis is aligned with the vertical direction.

## Data Availability

All data needed to evaluate the conclusions in the paper are present in the paper and/or the Supplementary Information. Additional data related to this paper may be requested from the authors

## Acknowledgements

This work was supported by the National Natural Science Foundation of China under grant numbers 12425206, 12272206, 92252204 and 12388101. The authors also acknowledge Prof. Chun-Xiao Xu, Chao Sun and Yihui Zhang at Tsinghua University for helpful discussions.

## Author contributions

**Z. B. Hou**: Framework design, Experiment, Simulation, Visulization & Writing; **J. D. Zhang**: Experiment, Simulation, Visulization & Writing; **Y. D. Li**: Framework design & Experiment; **Y. X. Jia**: Experiment, Supervision, Review & Editing; **W.-X. Huang**: Project Administration, Simulation, Supervision, Review & Editing.

## Competing interests

Authors declare that they have no competing interests.

# Supplementary Information

## Mass distribution measurements of real samaras

To measure the mass distribution of real samaras, we collected representative samples from various species and scanned their surfaces using a KSCAN-Magic Handheld 3D Scanner (0.02 mm accuracy). By assuming a uniform mass distribution, we calculated the COM for each sample and aligned the samara models such that the COM was positioned at the origin of a coordinate system aligned with the principal axes of inertia. We then identified the geometric center on the plane perpendicular to the axis with the largest moment of inertia, which typically corresponds closely to the samara's surface. Using the modeling files, we measured the maximum extents in two directions within this plane, defining these as the lengths of the longer and shorter edges. These measurements allowed us to calculate the relative COM location for each samara sample using Equation 1-1. The thickness distribution across different samaras is shown in Figure S6, and the corresponding parameters for each are listed in Table S1. Our results show that the mass distributions of the real samaras closely align with those observed in our samara-inspired framework for the corresponding flight modes.

## Theoretical model for periodic flight patterns

Among the samara-inspired flight modes discussed, as well as similar behaviors seen in natural samaras, there are two typical periodic modes, i.e. the AR and ST modes. To explore the self-sustaining mechanisms underlying these behaviors, a theoretical model is proposed. Considering the flattened shape of real samaras and the plate model used in our analysis, both reasonably satisfy the relationship $I_3 = I_1 + I_2$, where $I_1$ and $I_2$ represent the moments of inertia about the longer and shorter axes, respectively, and $I_3$ corresponds to the axis perpendicular to the plane of the samara.

Euler's equations of rotation in this context are:

$$I_1 \dot{\omega}_1 + I_1 \omega_2 \omega_3 = M_1 \tag{S1-1}$$

$$I_2 \dot{\omega}_1 - I_2 \omega_1 \omega_3 = M_2 \tag{S1-2}$$

$$(I_1 + I_2)\dot{\omega}_3 + (I_2 - I_1)\omega_2 \omega_3 = M_3 \tag{S1-3}$$

The angular velocities under a 3-2-1 Euler angle formulation are:

$$\omega_1 = \dot{\psi} + \dot{\varphi}\sin\theta \tag{S2-1}$$

$$\omega_2 = \dot{\varphi}\cos\theta\sin\psi \tag{S2-2}$$

$$\omega_3 = \dot{\varphi}\cos\theta\cos\psi \tag{S2-3}$$

Projecting torques onto the $y$-axis of the ground coordinate system yields:

$$M_y = M_2 \cos\psi - M_3 \sin\psi \tag{S3}$$

In the AR mode, the plate exhibits a steady, axisymmetric rotation during descent. The torque required to sustain rotation arises primarily from gravitational and aerodynamic forces. Based on kinematic measurements (Figure 3), we approximate that $\dot{\varphi}$ and $\theta$ remain constant while $\psi \approx 0$, yielding the following torque component along the $y$-axis:

$$M_y = -I_2 \dot{\varphi}^2 \sin\theta \cos\theta \tag{S4}$$

This torque acts in the negative $y$-direction. The AR motion promotes the formation of attached LEV and RLVs. These vortical structures induce a negative pressure zone near the plate's tip, generating lift that opposes gravity. Consequently, the aerodynamic lift ($f_z$), along with the gravity ($G$), generates torque, maintaining uniform rotation during descent (Figure S7A).

The ST mode is characterized by a spiral descent with rapid tumbling, which can be approximated as a precession with uniform angular velocity. Based on kinematic measurements (Figure 3), we assume $\dot{\psi}$, $\dot{\varphi}$, and $\theta$ remain constant, yielding the torque component along the $y$-axis:

$$M_y = -I_2(\dot{\varphi}\dot{\psi} + \dot{\varphi}^2 \sin\theta)\cos\theta \cos^2\psi + (I_1 + I_2)\dot{\varphi}\dot{\psi}\cos\theta \sin^2\psi \\ -(I_2 - I_1)\dot{\varphi}^2 \cos^2\theta \sin^2\psi \cos\psi \tag{S5}$$

By integrating Equation (S3) over a full tumbling period $T = \frac{2\pi}{<\dot{\psi}>}$, the torque required to sustain rotation, derived from gravitational and aerodynamic forces, yields an average $y$-component:

$$<M_y> = \frac{1}{T} \cdot \int_{t_0}^{t_0+T} M_y dt = I_1 \dot{\varphi}^2 \cos\theta \left(\frac{\dot{\psi}}{\dot{\varphi}} - \frac{I_2}{I_1}\sin\theta\right) \tag{S6}$$

For the continuous ST motion shown in Figure 3 (where $\frac{\dot{\psi}}{\dot{\varphi}} \approx 7.5$, $\frac{I_2}{I_1} \approx 8.9$, and $\theta \approx 38.2°$), the torque acts in the positive $y$-direction on average. This indicates that the combined effects of lift ($f_z$) and gravity ($G$) produce a torque (Figure S7B), guiding the plate along a spiral trajectory during continuous tumbling descent.

**Figure S1: Reconstruction of chaotic mode trajectory using experimental kinematic data.**

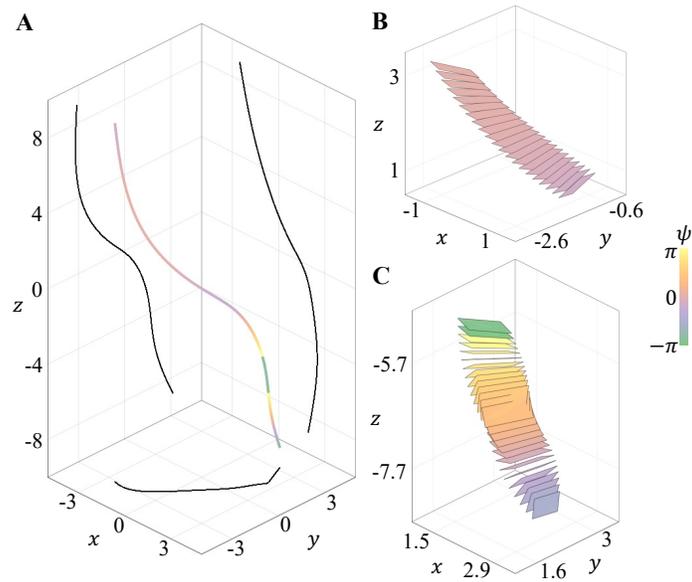

Trajectories of chaotic (CH) flight mode is visualized, with data normalized by the plate's length ($L$) and reconstructions overlaid at $5 ms$ intervals (see also movie S1). **(A)** Overall trajectory. **(B)** 3D reconstruction of the translational-dominated phase. **(C)** 3D reconstruction of the tumbling-dominated phase. 3D reconstructions use the self-rotation angle ($\psi$) around the plate's longer axis, coloring the plate from green to orange.

**Figure S2: Reconstruction of falling mode trajectory using experimental kinematic data.**

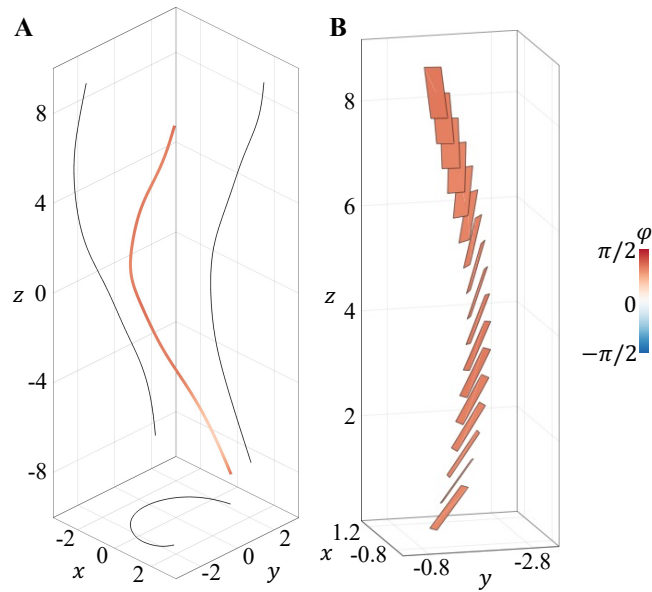

Trajectories of falling (FA) flight mode is visualized, with data normalized by the plate's length ($L$) and reconstructions overlaid at $15ms$ intervals (see also movie S1). **(A)** Overall trajectory. **(B)** 3D reconstruction of the falling flight mode. 3D reconstructions use the cone angle ($\theta$) between its longer axis and the horizontal plane, coloring the plate from blue to red.

**Figure S3: Schematic of experimental platform**

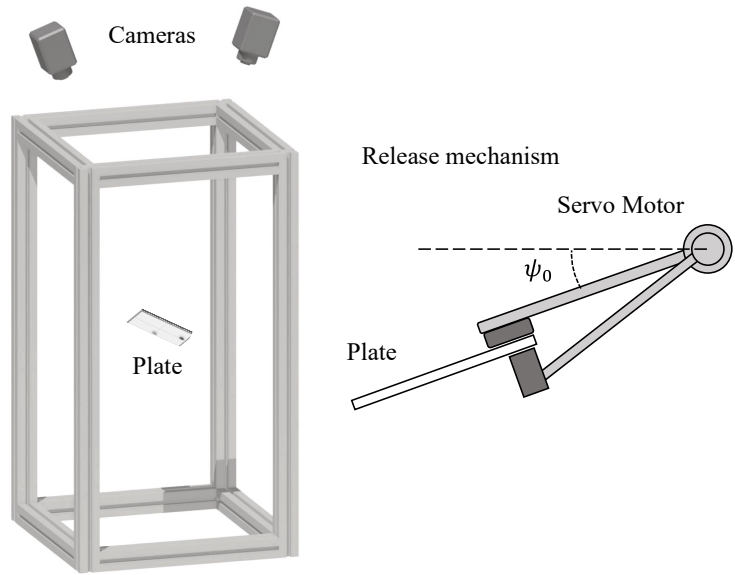

Experiment setup showing the overall arrangement with plate release mechanism.

**Figure S4: Simultaneous snapshots of the falling plate**

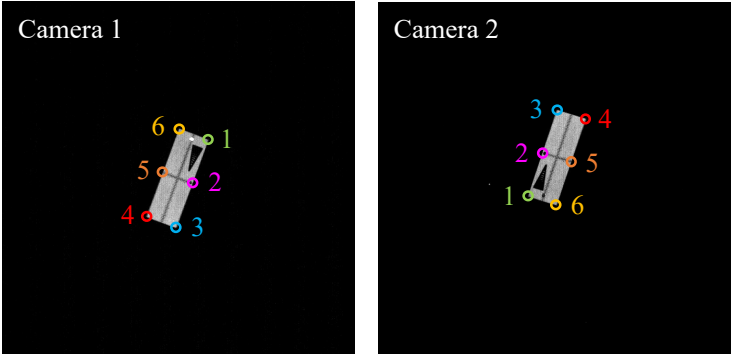

Scatters representing the six reference points on the plate used for kinematic measurements.

**Figure S5: Computational fluid dynamics simulation setup**

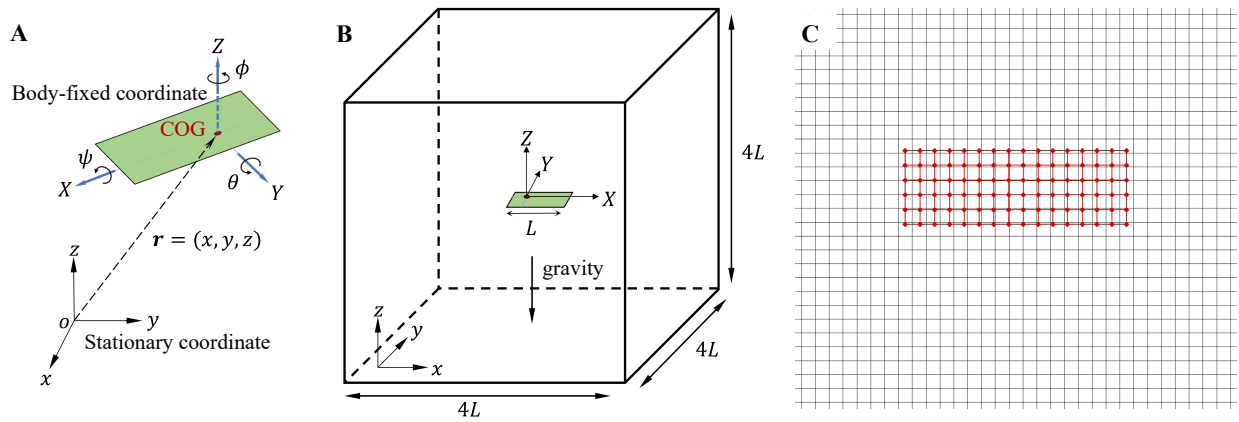

**(A)** Schematic of coordinate definition. **(B)** Numerical simulation domain. **(C)** Discretization in immersed boundary method: black mesh shows the background Eulerian grid; red mesh shows the Lagrangian grid on the plate.

**Figure S6: Mass distribution of real samaras**

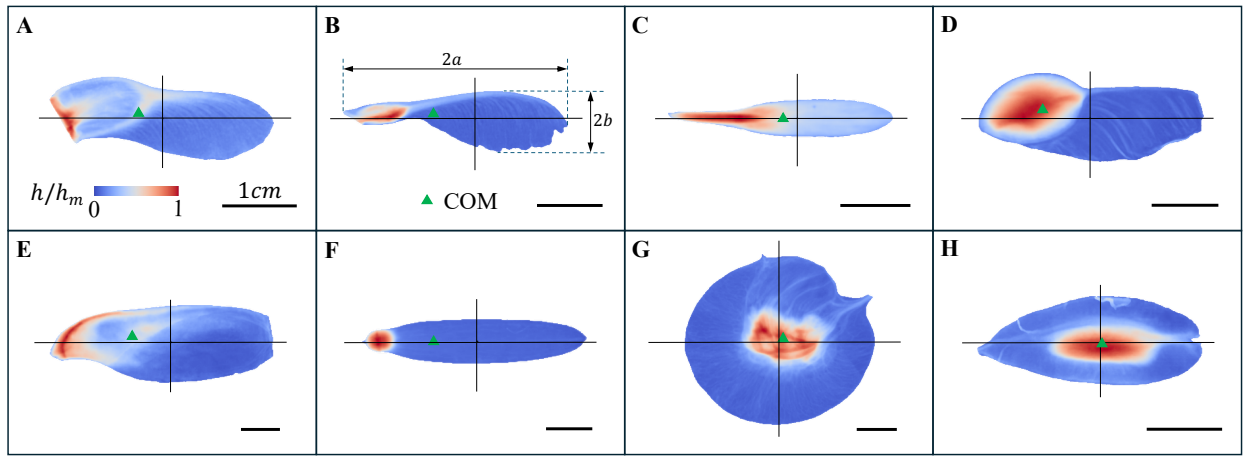

Normalized thickness distributions and center of mass (COM) locations are shown, based on measurements from actual samaras. **(A)** *Acer pictum* (maple), **(B)** *Acer negundo* (maple), **(C)** *Fraxinus* (ash), **(D)** *Pterolobium punctatum*, **(E)** *Swietenia mahagoni* (mahogany), **(F)** *Ventilago leiocarpa*, **(G)** *Pterocarpus santalinus* (red sandalwood), and **(H)** *Eucommia ulmoides*. The green triangles mark the COM locations, while the two orthogonal axes represent the directions parallel to the principal axes of rotational inertia, intersecting at the geometric center of each samara's plane. $h/h_m$ stands for the normalized thickness.

**Figure S7: Schematics of self-sustaining mechanism.**

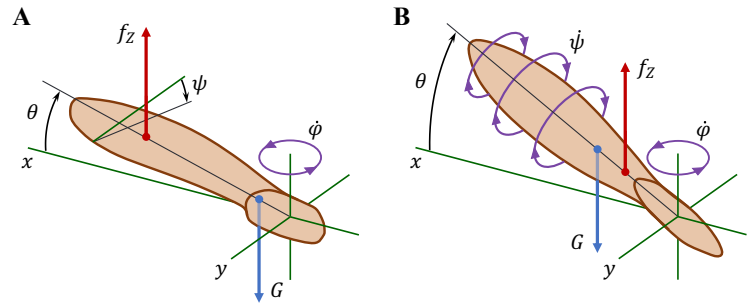

**(A)** AR mode. **(B)** ST mode. $\dot{\varphi}$ and $\dot{\psi}$ represent the angular velocities around the vertical axis and the plate's longer axis, respectively. Aerodynamic lift ($f_z$) and gravitational force ($G$) are also shown.

**Table S1 Mass distribution measurements of real samaras**

| Samaras | $x_c/a$ | $y_c/b$ | $2a$ (mm) | $2b$ (mm) | $h_m$ (mm) | $m_{total}$ (mg) | Mode |
|---|---|---|---|---|---|---|---|
| 1 | 0.21 | 0.12 | 30.12 | 11.20 | 1.78 | 50.7 | AR |
| 2 | 0.37 | 0.15 | 34.32 | 9.39 | 2.07 | 16.6 | AR |
| 3 | 0.12 | 0.01 | 32.09 | 5.45 | 1.82 | 33.7 | ST |
| 4 | 0.42 | 0.19 | 33.64 | 13.48 | 3.55 | 165.4 | AR |
| 5 | 0.35 | 0.15 | 57.93 | 20.09 | 2.96 | 53.8 | AR |
| 6 | 0.38 | 0.01 | 49.22 | 10.19 | 5.24 | 83.4 | FA |
| 7 | 0.04 | 0.04 | 47.82 | 43.37 | 4.27 | 503.9 | ST |
| 8 | 0.02 | 0.04 | 29.94 | 11.98 | 2.01 | 78.7 | ST |

This table summarizes the key parameters from the measurements of real samaras, including their morphological and mass distribution properties. $x_c/a$ and $y_c/b$ denote the relative coordinates of the COM along the $x$ and $y$ directions, respectively, with $2a$ and $2b$ representing the maximum extents in those directions within the samara plane. $h_m$ stands for the maximum thickness, and $m_{total}$ indicates the total mass of each samara. The corresponding flight modes are listed alongside these parameters. Images and detailed mass distribution of these samaras are shown in Figure 1A and Figure S6, respectively.

**Movie S1.**

3D reconstruction of flight modes in the samara-inspired framework

**Movie S2.**

Visualization of vortical structures in samara-inspired flight